# Statistical evaluation of 571 GaAs quantum point contact transistors showing the 0.7 anomaly in quantized conductance using millikelvin cryogenic on-chip multiplexing

Pengcheng Ma[1], Kaveh Delfanazari[1,2,3,*], Reuben K. Puddy[1], Jiahui Li[1], Moda Cao[1], Teng Yi[1], Jonathan P. Griffiths[1], Harvey E. Beere[1], David A. Ritchie[1], Michael J. Kelly[1,3] and Charles G. Smith[1]

[1]*Cavendish Laboratory, Department of Physics, University of Cambridge, Cambridge CB3 0HE, UK*

[2] *James Watt School of Engineering, University of Glasgow, Glasgow, Glasgow G12 8QQ, UK*

[3] *Electrical Engineering Division, University of Cambridge, Cambridge CB3 0FA, UK*

*Corresponding author: kaveh.delfanazari@glasgow.ac.uk, Dated: 25112023

The mass production and the practical number of cryogenic quantum devices producible in a single chip are limited to the number of electrical contact pads and wiring of the cryostat or dilution refrigerator. It is, therefore, beneficial to contrast the measurements of hundreds of devices fabricated in a single chip in one cooldown process to promote the scalability, integrability, reliability, and reproducibility of quantum devices and to save evaluation time, cost and energy. Here, we use a cryogenic on-chip multiplexer architecture and investigate the statistics of the 0.7 anomaly observed on the first three plateaus of the quantized conductance of semiconductor quantum point contact (QPC) transistors. Our single chips contain 256 split gate field effect QPC transistors (QFET) each, with two 16-branch multiplexed source-drain and gate pads, allowing individual transistors to be selected, addressed and controlled through an electrostatic gate voltage process. A total of 1280 quantum transistors with nano-scale dimensions are patterned in 5 different chips of GaAs heterostructures. From the measurements of 571 functioning QPCs taken at temperatures $T$= 1.4 K and $T$= 40 mK, it is found that the spontaneous polarisation model and Kondo effect do not fit our results. Furthermore, some of the features in our data largely agreed with van Hove model with short-range interactions. Our approach provides further insight into the quantum mechanical properties and microscopic origin of the 0.7 anomaly in QPCs, paving the way for the development of semiconducting quantum circuits and integrated cryogenic electronics, for scalable quantum logic control, readout, synthesis, and processing applications.

*Introduction*

Semiconductor field effect mesoscopic devices, controlled and addressed by gate voltages, are fundamental in quantum science and technology applications. Quantum mechanics dominates their electronic properties and responses at cryogenic temperatures. The study of device characteristics often is based on the measurements of a single device, however even devices with an identical design and having the same fabrication process result in different physical characteristic properties. In this category, gate-defined semiconducting quantum point contact (QPC) transistors[1] or split-gate QPC field effect transistors (QFET) are necessary cryogenic quantum nano-electronic devices in spin and charge qubit-based quantum processors. Their differential conductance is quantized in units of $G_Q=2e^2/h$, where $e$ is the electron charge, and $h$ is Planck's constant [2,3]. However, there is an unanticipated shoulder at ~0.7 $G_Q$, referred to as the '0.7 anomaly' [4,5], which is still not fully understood despite twenty years of research[6,7]. Various models have been proposed, including spontaneous spin polarisation[4,8-10], Wigner crystallisation[11,12], the Kondo effect[13-16,] and the smeared van Hove singularity[17-20]. Similarly, several experiments have been performed. For instance, Al-Taie *et al.* used an array of QPCs, with electron motion in the channels in the *x*-direction and the lateral confinement in the *y*-direction, and performed statistical quantum transport measurements at low temperatures[21-23]. They expressed the barrier and confinement curvatures in terms of the harmonic-oscillator energies, $E_x = \hbar\omega_x$ and $E_y = \hbar\omega_y$, and studied the effect of disorders on quantum device performance. Moreover, Bauer *et al.* used the combination of a potential barrier and electron interactions to modify the van Hove singularity as a ridge-like structure in the local density of states (*LDOS*)[17].

*Theory of QPC transistors*

In this work, by studying a large array of on-chip integrated QPC devices we show that some of the features in our experimental results, taken at both $T=40$ mK and $T=1.4$ K, largely support the van Hove model with short-range interactions. The *LDOS* is a function of the dimensionless quantity $\kappa=(\mu-V_c)/E_x=\alpha e(V_G-V_G^{\text{riser}})/E_x$, where $\mu$ is the chemical potential, $V_c$ is the central barrier height, $\alpha=dV_{SD}/dV_{SG}$ is the lever arm, $V_G$ is the split gate voltage and $V_G^{\text{riser}}$ is the split-gate voltage at the plateau riser. Electrons are slowed down where the *LDOS* is high and experience an enhanced effective interaction strength $U_{\text{eff}}(\kappa)=U \cdot LDOS(\kappa)$. Here, $\kappa$ is proportional to $V_G$ through $\kappa=\alpha e(V_G-V_G^{\text{riser}})/E_x$ and $U$ is the interaction strength as defined in *Eq.* A1 of the Appendix. From the expression $1-U_{\text{eff}}= dV_c^h/dV_c$ [19], where $V_c^h$ is the effective Hartree barrier height, as described in *Eq.* A2 in the Appendix, we can deduce

$$1-U \cdot LDOS(\kappa) = 1-U_{eff}(\kappa) \approx S_{TC}(\kappa)$$

(1)

Here, transconductance suppression $S_{TC}$ is defined as the ratio between the measured transconductance ($TC_{SD}=dG_{SD}/d\kappa$ and $G_{SD}$ is the measured source-drain conductance) and non-interacting transconductance ($TC^0=dG^0/d\kappa$ and $G^0$ is the non-interacting conductance). The non-interacting variables are marked with a superscript 0 throughout the paper.

Equation (1) implies a direct link between $S_{TC}$ and the *LDOS* ridge. The maximum of the *LDOS* ($LDOS_{max}$) results in the strongest effective interaction strength ($U_{eff}^{max}$), and thus the strongest transconductance suppression. We define the minimum of transconductance suppression as the '0.7 anomaly transconductance suppression $S_{TC}^{0.7}$'. The position $\kappa$ and conductance $G$ at the minimum of transconductance suppression are defined as $\kappa_{TC}^{0.7}$ and $G_{TC}^{0.7}$. Therefore,

$$1 - U \cdot LDOS_{max} = 1 - U_{eff}^{max} \approx S_{TC}^{0.7}$$

(2)

This equation shows the link between $S_{TC}^{0.7}$ and $U_{eff}^{max}$. The $LDOS_{max}$ and $U$ scale with $1/\sqrt{E_x}$ [17] and $\sqrt{E_y}$ [24], respectively. Substitution of these expressions into *Eq.* (2) gives

$$U_{eff}^{max} \propto \sqrt{E_y/E_x}$$

(3)

To explore the role of $U_{eff}^{max}$, it is necessary to investigate the statistics of $E_x$ and $E_y$ in a large array of devices and study the geometry effect on them (we define $E_y/E_x=U_E$ to characterise $U_{eff}^{max}$).

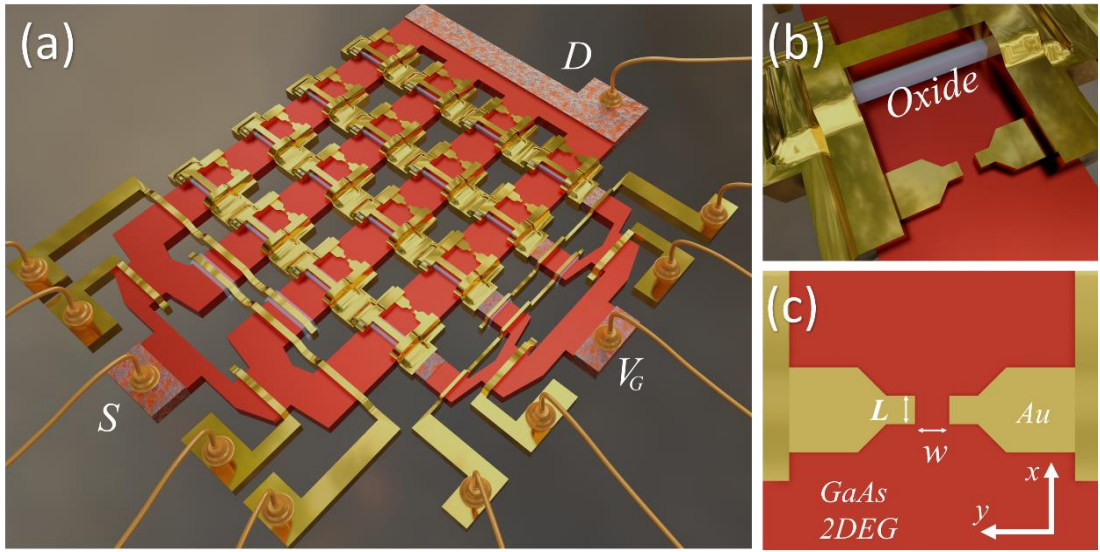

Figure 1 **3D schematic of the GaAs multiplexed quantum chip. a**, Overall structure of a 4-by-4 MUX design with Au gate represented by the bright metal yellow, and rough orange for Au/Ge/Ni Ohmic alloy. The smooth red connected to the ohmic metal is the channel formed by the MESA wet etch, and it is a 90nm thick GaAs 2DEG buried underneath. '$S$', '$V_G$' and '$D$' are labelled on the MESA represent source, drain and gate pads, respectively. The transparent blue under the gates is the metal gates insulating layer (oxide). Finally, the bright orange metal is the wire bonding connected to the pads. **b**, The enlarged view of a single addressable device in the 4-by-4 MUX array in split-gates representations. **c**, The plane top view of the split gates with the direction of $E_x$ and $E_y$ indicated.

*Fabrication of large arrays of QPCs with on-chip multiplexing architecture*

With this aim, we developed a cryogenic multiplexer (MUX) architecture for large-scale nanofabrication and on-chip integration of QPC transistors. A total of 1280 quantum nanodevices on five different chips (labelled samples 1-5) are designed and manufactured. Each single chip contains 256 split-gate QPC transistors, with two 16-branch multiplexed source-drain and gate pads. Through this cryogenic multiplexed architecture, individual transistors are selected, addressed and controlled by using the electrostatic gating method. The wafer used for this experiment is a GaAs/AlGaAs heterostructure. First, optical lithography is performed followed by a 100nm deep MESA chemical wet etch with $H_2SO_4: H_2O_2: H_2O = 1: 8: 1600$ for 150s. Secondly, the Ohmic contacts are defined with another optical lithography with an 'undercut' profile. The ohmic metal (AuGeNi alloy) is annealed at 430℃ for 80s so that the metal diffuses into the 2DEG and makes contact with the MESA pattern from the previous step. An insulator is required for gates on the device.

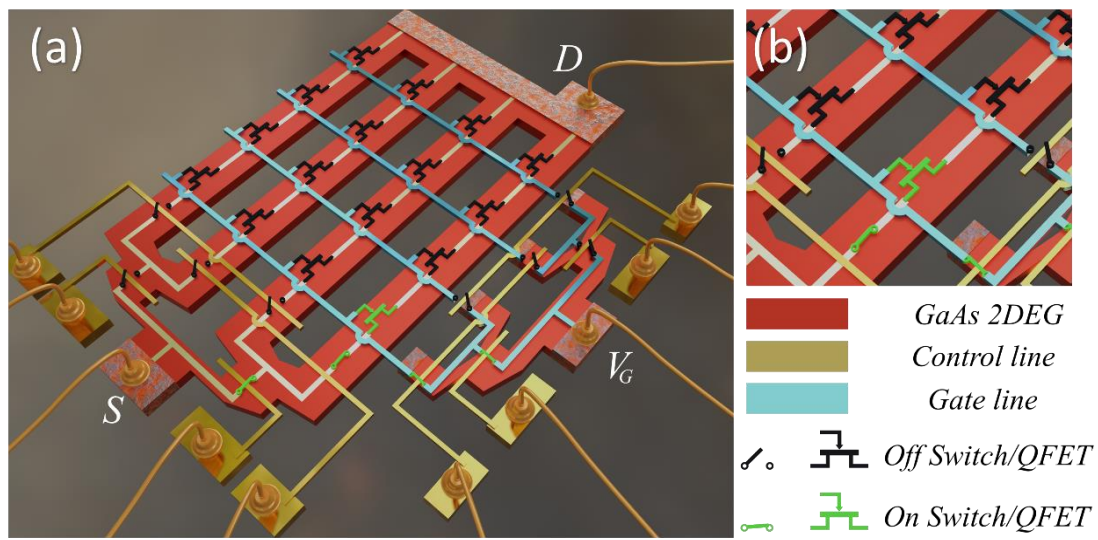

Figure 2 **Circuit representation of the GaAs field effect multiplexed quantum chip. a**, Circuit equivalent of a 4-by-4 MUX quantum field effect transistors with gate line represented by the bright blue, and control lines for both channel and gates are bright metal yellow. The white line represents the device conduction path. The off-state QFET and switch are black and the ones that are on are bright green. **b**, The enlarged view of the opened device with green switches switched on to activate that single QFET device.

A mixture of polyimide and T9039 with a 2:1 ratio is used as the insulator between the gates and the sample. After the insulating layer deposition, gates are defined with a similar process as the ohmic metal but with Ebeam-lithography as the QPC requires defining fine pattern. The final step is wire bonding and mounting the sample to the LCC chip carrier followed by loading in the cryostat or dilution fridges for cryogenic characterisation. A reduced-number MUX 3D model is shown in Fig. 1 along with the circuit representation in Fig. 2. Figure 1a explicitly shows the structure of a 4-by-4 MUX design with red 2DEG and bright gold gates on top. A single split-gate QFET is shown in the enlarged view of Figure 1b. Figure 2a is the circuit equivalent to the same MUX design shown in Fig. 1a. A zoomed view also shows how a single functioning

QFET is addressed by control lines which are represented by the bright green switches in Fig. 2b. Our MUX chips contain 256 QPCs of different widths $W$ and lengths $L$, allowing the study of the roles of $E_x$ and $E_y$ (the shape of the saddle point potential) on the 0.7 anomaly. From low-temperature quantum transport measurements, we find that $S_{TC}$ is strongest for the first plateau and stronger at the higher temperature of $T$=1.4 K compared to $T$=40 mK. A very strong $S_{TC}$ can result in the spontaneous risers-splitting of transconductance (i.e. the appearance of two peaks on the transconductance curve between adjacent subbands).

Figure 3a shows a Scanning Electron Microscope (SEM) image of our semiconducting quantum multiplexed chip consisting of an array of 256 QPC transistors addressed by two MUXs. The selected area in Fig. 3a is enlarged in Fig. 3b, showing three QPCs in a row. Figure 3c shows a QPC with width $W$ and length $L$ labelled in the image.

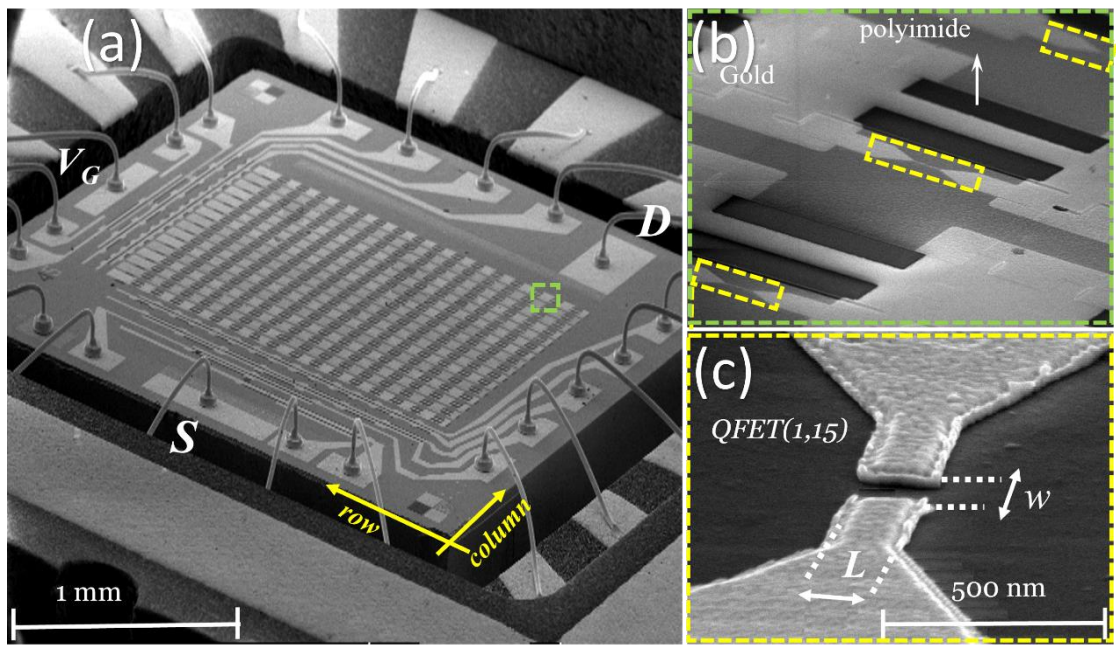

Figure 3 **SEM images of one GaAs quantum chip with a large array of multiplexed QFETs. a**, A quantum multiplexed (MUX) circuit consisting of 256 QPCs (QFETs) that are controlled and read out by only 19 electrical contacts. Each QPC is measured by addressing the row of source-drain multiplexer ($S$ and $D$ contacts are labelled in the figure) and the column of gates MUX ($V_G$ contact is labelled in the figure), and is named in terms of the coordinate QFET (row, column). **b**, The enlarged area of the dashed-square defined in **a**, showing three split-gate quantum transistors in the first row QFET (1,14), QFET (1,15) and QFET (1,16) (marked with three dashed-rectangles). Polyimide is shown in black. **c**, The device QFET (1,15) with width $W$ and length $L$ labelled.

*Cryogenic characterisation of on-chip multiplexed QPC transistors*

In total, 571 QPCs are successfully measured in chips 1-5 (out of a total of 1280 QPCs). First, only devices that show conductance suppression are investigated with a focus on the 0.7 anomaly. Second, only devices that have a good fitting of $G^0$ with $G_{SD}$ are investigated to ensure that the device has a parabolic potential barrier, as a precondition of the van Hove formulas. Here, a good fit is where $G^0$ well coincides with $G_{SD}$. To extract $E_x$, we fit the lower half step of $G^0$ with that of corrected $G_{SD}$ based on the Landauer-Buttiker formula[25]. We extract $E_x^N$ in these devices for the first three subbands ($N$ is the subband index), in two forward (as the conductance decreases with gate voltage) and backward (as the conductance increases with gate voltage) sweep directions. Figure 4a shows a scatter plot of $E_x$ values between the first ($x$-axis) and third ($y$-axis) cool-downs for $n= 33$ devices in sample 1 at $T=40$ mK. $E_x$ values are distinct for a different cooldown (from room temperature down to $T=40$ mK), indicating that $E_x$ is highly influenced by random fluctuations of the electrostatic potential from the different cooldowns. Figure 4b shows the geometry dependence of $E_x$ for $n= 62$ devices in sample 1 at $T=40$ mK. $E_x$ values are nearly independent of the device length. In particular, the $E_x$ values vary a lot for samples of the same length. The red dots (devices with width $W_1=0.6$ μm) and blue crosses (devices with width $W_2=0.4$ μm) are not distinct, showing $E_x$ values are also independent of the device width. This suggests $E_x$ is more influenced by the potential background than device geometry.

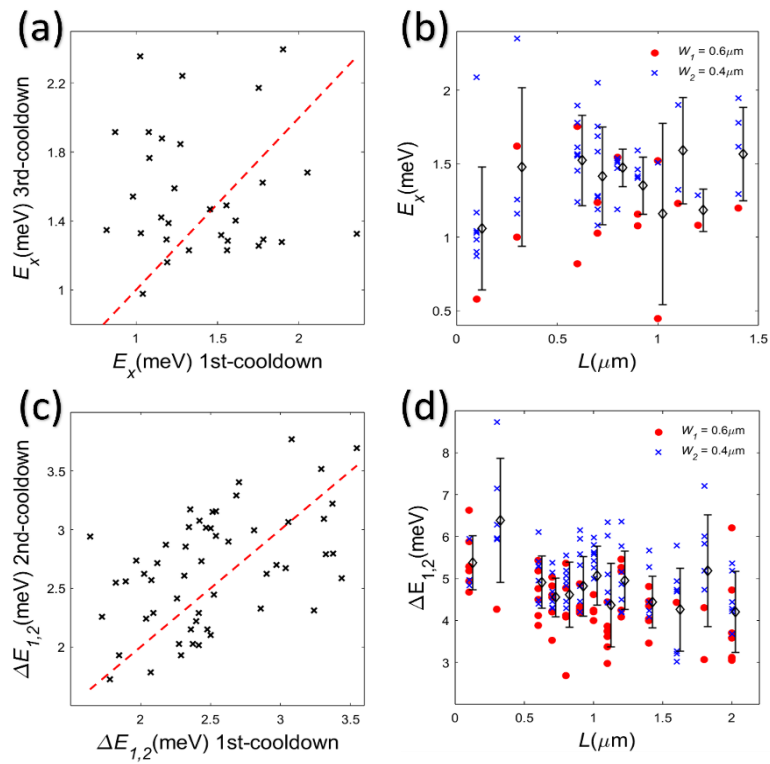

Figure 4 **Cryogenic evaluation of large array multiplexed QPCs. a**, Geometry dependence of the potential curvatures. Scatter plot of $E_x$ between the first ($x$-axis) and third ($y$-axis) cooldowns for sample 1 at $T=40$ mK. $n$ counts the measured device number. The red dashed line (gradient=1) provides a guide to the eye. The results are in the forward sweep if not specified. **b,** Geometry dependence of $E_x$ for sample 1 at $T=40$ mK. The red dots and blue crosses represent devices with widths $W_1=0.6$ μm and $W_2=0.4$ μm, respectively. The error bar represents the mean ± standard deviation for devices at each length, offset horizontally by 0.025 μm for clarity. **c**, Scatter plot of $\Delta E_{1,2}$ between the first ($x$-axis) and second ($y$-axis) cooldowns for sample 1 at $T=40$ mK. **d**, Geometry dependence of $\Delta E_{1,2}$ for sample 1 at $T=1.4$ K after illumination.

The values of $E_y$ are equal to the subband spacings $\Delta E_{N,N+1}$ for a parabolic barrier and are extracted using DC bias spectroscopy. The drop of the applied DC bias voltage across the series resistance is corrected by using *Eq*. A3 in the Appendix. Figure 4c is a scatter plot of $\Delta E_{1,2}$ values between the first and second cooldowns for *n*= 54 devices in sample 1 at *T*=40 mK, showing a repeatable correlation between the two measurements. The values of $\Delta E_{1,2}$ are greatly amplified after illumination, presumably because illumination increases the carrier concentration so that a larger split-gate voltage is required to pinch off the device leading to a greater confining potential. Figure 4d shows the geometry dependence of $\Delta E_{1,2}$ for *n*= 143 QFET nanodevices in sample 1 at *T*=1.4 K. For length dependence, the means of $\Delta E_{1,2}$ (marked with diamonds) show a weak negative correlation with length. For width dependence, the red dots tend to be below the blue crosses, indicating a larger width can result in smaller values of $\Delta E_{1,2}$. Thus, $\Delta E_{1,2}$ is weakly correlated with both the length and width, which accords with the modelling by Koop *et al.*[26].

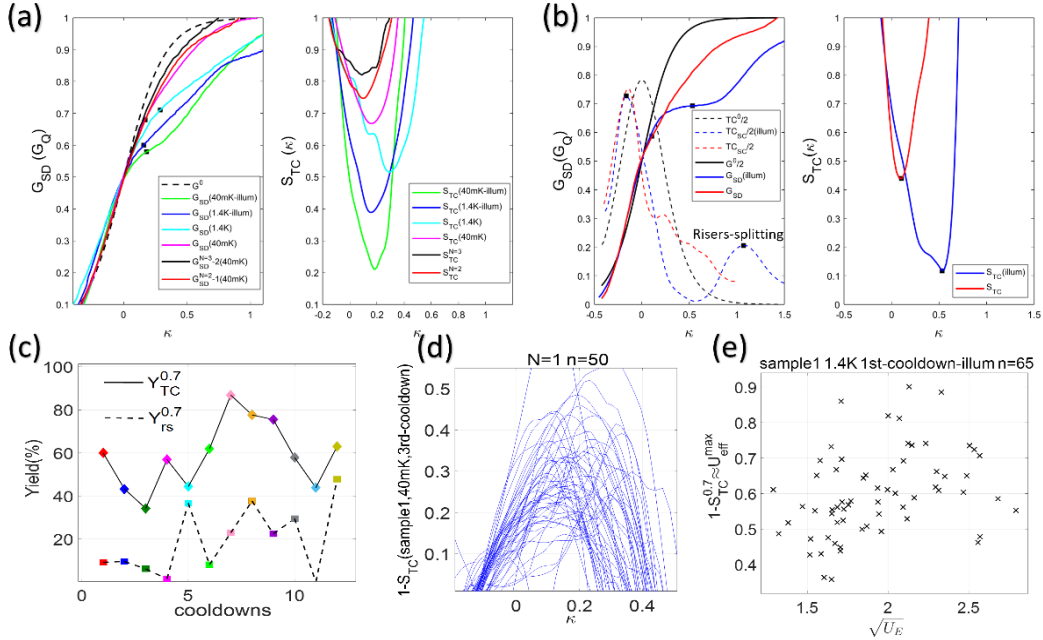

Figure 5 **Transconductance suppression of the 0.7 anomaly in large array multiplexed QPCs.** **a**, Comparisons of $S_{TC}(\kappa)$ (right) at *T*=40 mK and 1.4 K before and after illumination, respectively, for device *D* (16,15), sample 1. **b**, Spontaneous risers-splitting for transconductance for device *D* (10,11), sample 1 at *T*=40 mK. **c**, Comparisons of $Y_{rs}^{0.7}$ (dashed line) with $Y_{TC}^{0.7}$ (solid line) for samples 1-5 in 12 cooldowns. $Y_{rs}^{0.7}$ ( $Y_{TC}^{0.7}$ ) is the yield of 1D devices showing risers-splitting (transconductance suppression). **d**, Curves of 1-$S_{TC}$ ($\approx U \cdot LDOS$) as a function of $\kappa$ for the first plateau, for sample 1 at *T*=40 mK. *n* counts the device number. **e**, Correlation of 1-$S_{TC}^{0.7}$ with $\sqrt{U_E}$ in *n*= 65 devices, in sample 1 at *T*=1.4 K, after illumination. $\rho$(1-$S_{TC}^{0.7}$, $\sqrt{U_E}$)= 0.301.

Therefore, unlike $E_x$, which is difficult to adjust, $E_y$ can be adjusted using the device geometry. A short and narrow QPC yields the strongest confinement. In this way, the control of $U_{eff}^{max}$ can be realised. Figure 5a compares the $S_{TC}$ of device *D* (16,15), sample 1 at $T$=1.4 K, and $T$=40 mK before and after illumination. The device has a width $W$=0.4 μm and a length $L$=0.1 μm. $V_G$ axis is transformed into $\kappa$ by $\kappa=\alpha e(V_G-V_G^{riser})/E_x$, where $\alpha$ is the lever arm and $V_G^{riser}$ is the split-gate voltage at the plateau riser with a conductance of 0.5 $G_Q$. The $\kappa_{TC}^{0.7}$ position is marked with black squares for each $G_{SD}$ curve. $S_{TC}$ is reduced (read the right axis) after illumination at both $T$=40 mK and $T$=1.4 K. For higher plateaus, $G_{SD}^{N=2}$ (yellow curve) and $G_{SD}^{N=3}$ (brown curve) are vertically offset by $G_Q$ and 2 $G_Q$, respectively, to the first plateau position. $S_{TC}^{N=2}$ and $S_{TC}^{N=3}$ are less reduced than $S_{TC}$ for the first plateau.

The key points are as follows: (i), $S_{TC}$ values are smaller at $T$=1.4 K than those at $T$=40 mK. This effect is consistent with the temperature dependence of the 0.7 anomaly as commonly reported[4] (i.e. the 0.7 anomaly becomes more pronounced as temperature increases[1]). (ii), $S_{TC}$ values become smaller after illumination at both $T$=1.4 K and $T$=40 mK temperatures since illumination increases $E_y$ but not $E_x$, resulting in larger values of $U_{eff}^{max}$. (iii), for higher plateaus, $S_{TC}^{N=3}$< $S_{TC}^{N=2}$< $S_{TC}^{N=1}$. This is because when the 1D subband index increases, the width increases, and the density in each sub-open regime increases [17] (i.e. at 0.5-0.9 $G_Q$), and $U_{eff}^{max}$ drops to almost zero for the third plateau. Figure 5b illustrates the spontaneous riser-splitting of the transconductance for device $D$ (10,11), sample 1 at $T$=40 mK. $TC_{SD}$ curve is the differential of smoothed $G_{SD}$. For $TC^0$, $TC_{SD}$ and $S_{TC}=TC_{SD}/TC^0$ curves, read the right axis. $S_{TC}^{0.7}$ is marked with triangles on $S_{TC}$ curves before and after illumination. After illumination, due to a strong suppression of $S_{TC}$ (circle-marked blue line), two split-risers (marked with dots) appear on $TC_{SD}$ curve (star-marked green line).

After illumination, $TC_{SD}$ is suppressed and $S_{TC}^{0.7}$ decreased from a value of 0.44 before illumination to a value of 0.12, which gives rise to a local minimum and leads to the riser-splitting as indicated by the arrow in the $TC_{SD}$ curve (star-marked green line). Thomas *et al*. regarded the riser-splitting as evidence for the spin-gap of spontaneously polarised spins[5]. However, after doing a statistical check on numerous devices, we find that the riser-splitting has a very low probability of occurrence, whereas the 0.7 anomaly suppression is observed much more often. Figure 5c compares the yield of 1D devices showing the riser-splitting $Y_{rs}^{0.7}$ to the yield of devices showing transconductance suppression $Y_{TC}^{0.7}$ in samples 1-5 in 12 different cooldowns. It is found that the number for $Y_{rs}^{0.7}$ is much lower than the number for $Y_{TC}^{0.7}$ at each cooldown. It is therefore inferred that the spontaneous polarisation model may not fit our results. A very large $U_E=E_y/E_x$, and the suppression of transconductance, but not spontaneous spin polarisation, cause the riser-splitting. Based on *Eq.* (1), we can use the expression $1-S_{TC}(\kappa)\approx U\cdot LDOS(\kappa)$ to roughly characterise the effective $LDOS(\kappa)$ shape. Figure 5d shows curves of $1-S_{TC}$ for the first plateau, for sample 1 at $T$=40 mK. The number of devices showing $S_{TC}$ decreases with increasing $N$, showing a decreasing chance of the 0.7 anomaly for higher subbands. Furthermore, Fig. 5e shows that in sample 1 at $T$=1.4 K, the Pearson correlation coefficient $\rho$ of $1-S_{TC}^{0.7}$ with $\sqrt{U_E}$, shown as $\rho(1-S_{TC}^{0.7}, \sqrt{U_E})$= 0.301, indicating $1-S_{TC}^{0.7}$ is positively correlated with $\sqrt{U_E}$. This obeys expression $1 - S_{TC}^{0.7} \propto \sqrt{U_E}$, which is a combination of *Eq.* (2) and (3). Note that $E_x$ also has an effect on $S_{TC}^0$ in the non-interacting regime.

Next, we discuss the interaction effects on the whole curve of conductance suppression $S_G = G_{SD}/G^0$ in our multiplexed quantum circuits. Figure 6a shows the first step of conductance for $n= 62$ devices in sample 1 at $T=40$ mK. The lower half step assembles together, and the 0.7 anomaly manifests different suppression levels compared with $G^0$ in the sub-open regime. The $E_x$ and $1/U_E$ dependence of $S_G$ at fixed $\kappa$ positions (see vertical coloured lines) are shown in Fig. 6b and Fig. 6c, respectively. It can be seen that $S_G$ is positively correlated with $E_x$ and $1/U_E$. Figure 6d compares the Pearson correlation coefficient $\rho$ of $S_G$ with $E_x$, shown as $\rho(S_G, E_x)$, indicated with dots, and $1/U_E$, shown as $\rho(S_G, 1/U_E)$, indicated with diamonds, for different cooldowns. The diamond-marked lines are above the dot-marked lines, indicating $\rho(S_G, 1/U_E) > \rho(S_G, E_x)$, and $S_G$ is more correlated with $1/U_E = E_x/E_y$ than $E_x$. This underlines that the role of $E_y$ on the 0.7 anomaly should be considered as being the same as that of $E_x$.

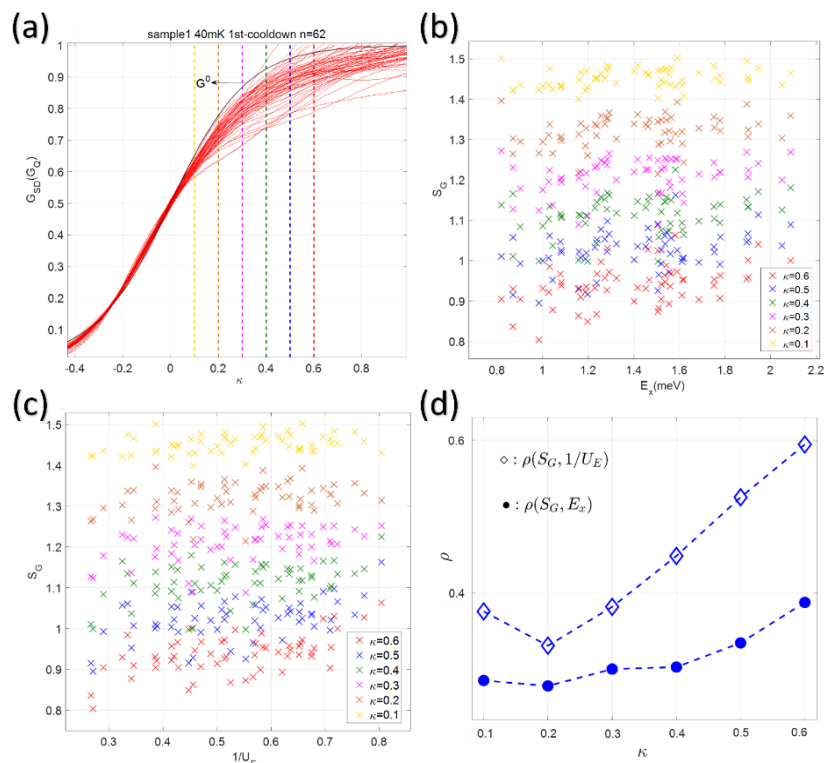

**Figure 6 Interaction effects on conductance suppression at fixed $\kappa$ positions in large array multiplexed QPCs.** **a**, The first step of quantized conductance for $n= 62$ QPC devices in sample 1 at $T=40$ mK. **b,c**, Dependence of $S_G$ (offset upward in turn by 0.1 for clarity) on $E_x$ (**b**) and $1/U_E$ (**c**) at fixed $\kappa$ positions (vertical coloured lines in **a**). **d**, Comparisons of $\rho(S_G, 1/U_E)$ (marked with diamonds) with $\rho(S_G, E_x)$ (marked with dots) as a function of $\kappa$.

*Summary*

In short, we developed a cryogenic chip-integrated multiplexed quantum electronic circuit and demonstrated the statistical measurements of 571 split-gate quantum wires at sub-Kelvin temperature. From the measurements taken at temperatures $T= 1.4$ K and 40 mK, we found that our data largely agreed with van Hove model with short-range interactions. The 0.7 anomaly shows the strongest suppression of transconductance, which is governed by the ratio of the saddle point potential curvatures $E_y/E_x$. Moreover, we realised that the device geometry can be made to influence $E_y$ but not $E_x$. The latter is more sensitive to and is dominated by the potential background and is the same for short or long barriers of a given width of the channel, while the former is more geometry-dependent. We highlight that our method is a generic platform and can be applied to most kinds of material systems for further development of quantum devices and integrated circuits[24-30]. For instance, implementation of the presented integrated cryogenic on-chip multiplexing architecture to hybrid superconducting-semiconducting junctions may open up an avenue towards a better understanding of interfacial and geometrical effects in the behaviour of hybrid junctions, for topological superconducting network[31-41]. The proposed architecture can also be used for the control and readout of hybrid quantum electronic circuits for their potential applications in cryogenic nanoelectronics and fault-tolerant quantum processing.

*Materials and methods*

Samples 1-5 are fabricated on a modulation-doped GaAs/AlGaAs heterostructure with two-dimensional electron gas (2DEG) formed 90 nm below the surface. The split gates are defined by electron-beam lithography. For sample 1, the carrier density and mobility are 1.71 (3.39)$\times 10^{11}$ cm$^{-2}$ and 1.59 (3.82)$\times 10^{6}$ cm$^{2}$V$^{-1}$s$^{-1}$ before (after) illumination, respectively. For samples 2-5, the carrier density and mobility are 2.08 (3.06)$\times 10^{11}$ cm$^{-2}$ and 3.06 (5.15)$\times 10^{6}$ cm$^{2}$V$^{-1}$s$^{-1}$ before (after) illumination, respectively. Samples 1, and 2 have the QFET device geometry with length $L$ varying at fixed width $W_1$=0.6 μm or $W_2$=0.4 μm. Samples 3-5 have the device geometry with a fixed geometrical aspect ratio $L/W$. We perform two-terminal measurements using standard lock-in and AC excitation at 77 Hz. Samples 1-5 are measured at $T$=1.4 K in a $^4$He cryostat. Sample 1 is measured at $T$=40 mK in a dilution refrigerator in the absence or presence of magnetic fields. We use a red light-emitting diode (LED) for illumination purposes. The quantum transport measurements of the large arrays of chip-integrated multiplexed quantum devices were performed fully automated and performed by MATLAB.

*Acknowledgements*

The authors acknowledge financial support from EPSRC, UK. P. C. Ma would like to thank the China Scholarship Council (CSC) for its financial support. The data presented in this paper can be accessed at https://doi.org/xx.xxxxx/CAM.xxxxx.

*References*

[1] Thornton, T. J. *et al*. One-dimensional conduction in the 2D electron gas of a GaAs-AlGaAs heterojunction. *Phys. Rev. Lett*. **56**(11):1198-1201 (1986)

[2] van Wees, B. J. *et al*. Quantised conductance of point contacts in a two-dimensional electron gas. *Phys. Rev. Lett.* **60**, 848-850 (1988).

[3] Wharam, D. A. *et al*. One-dimensional transport and the quantisation of the ballistic resistance. *J. Phys. C* **21**, L209-L214 (1988).

[4] Thomas, K. J. *et al*. Possible spin polarisation in a one-dimensional electron gas. *Phys. Rev. Lett.* **77**, 135-138 (1996).

[5] Thomas, K. J. *et al*. Interaction effects in a one-dimensional constriction. *Phys. Rev. B* **58**, 4846-4852 (1998).

[6] Micolich, A. P. What lurks below the last plateau: experimental studies of the 0.7×2e$^2$/h conductance anomaly in one-dimensional systems. *J. Phys. Condens. Matter* **23**, 443201 (2011).

[7] Micolich, A. P. Quantum point contacts: Double or nothing? *Nat. Phys.* **9**, 530-531 (2013).

[8] Wang, C. K. and Berggren K. F., Spin splitting of subbands in quasi-one-dimensional electron quantum channels. *Phys. Rev. B* **54**, R14257-R14260 (1996).

[9] Bruus, H. *et al.* The anomalous 0.5 and 0.7 conductance plateau in quantum point contacts. *Physica E* **10**, 97-102 (2001).

[10] Reilly, D. J. *et al.* Density-dependent spin polarisation in ultra-low-disorder quantum wires. *Phys. Rev. Lett.* **89**, 246801 (2002).

[11] Matveev, K. A. Conductance of a quantum wire in the Wigner-crystal regime. *Phys. Rev. Lett.* **92**, 106801 (2004).

[12] Brun, B. *et al.* Wigner and Kondo physics in quantum point contacts revealed by scanning gate microscopy. *Nat. Commun.* **5**, 4290 (2014).

[13] Cronenwett, S. M. *et al*. Low-temperature fate of the 0.7 structure in a point contact: a Kondo-like correlated state in an open system. *Phys. Rev. Lett.* **88**, 226805 (2002).

[14] Meir, Y *et al.* Kondo model for the 0.7 anomaly in transport through a quantum point contact. *Phys. Rev. Lett.* **89**, 196802 (2002).

[15] Rejec, T. and Meir, Y. Magnetic impurity formation in quantum point contacts. *Nature* **442**, 900-903 (2006).

[16] Iqbal, M. J. *et al.* Odd and even Kondo effects from emergent localisation in quantum point contact. *Nature* **501**, 79-83 (2013).

[17] Bauer, F. *et al*. Microscopic origin of the '0.7-anomaly' in quantum point contacts. *Nature* **501**, 73-78 (2013).

[18] Bauer, F. *et al.* Functional renormalisation group approach for inhomogeneous interacting Fermi systems. *Phys. Rev. B* **89**, 045128 (2014).

[19] Heyder, J. The 0.7 anomaly in quantum point contacts: a microscopic model for the first conductance step. *PhD Thesis* (2014).

[20] Heyder, J. *et al.* Relation between the 0.7 anomaly and the Kondo effect: Geometric crossover between a quantum point contact and a Kondo quantum dot. *Phys. Rev. B* **92**, 195401 (2015).

[21] Al-Taie, H. *et al.* Cryogenic on-chip multiplexer for the study of quantum transport in 256 split-gate devices. *Appl. Phys. Lett.* **102**, 243102 (2013).

[22] Smith L. W. *et al*. Dependence of the 0.7 anomaly on the curvature of the potential barrier in quantum wires. *Phys. Rev. B* **91**, 235402 (2015).

[23] Smith L. W. *et al*. Effect of split gate size on the electrostatic potential and 0.7 anomaly within quantum wires on a modulation-doped GaAs/AlGaAs heterostructure. *Phys. Rev. Applied* **5**, 044015 (2016).

[24] Lunde, A. M. *et al*. Electron-electron interaction effects in quantum point contacts. *New J. Phys.* **11**, 023031 (2009).

[25] Buttiker, M. Quantized transmission of a saddle-point constriction. *Phys. Rev. B* **41**, 7906(R) (1990).

[26] Koop, E. *et al*. The influence of device geometry on many-body effects in quantum point contacts: signatures of the 0.7-anomaly, exchange and Kondo. *J. Supercond. Nov. Magn.* **20**, 433-441 (2007).

[27] Witt J. D. S. *et al.* Spin-relaxation mechanisms in InAs quantum well heterostructures, *Appl. Phys. Lett*. 122, 083101 (2023)

[28] Puddy R. K. *et al.* Multiplexed charge-locking device for large arrays of quantum devices, *Appl. Phys. Lett.* 107, 143501 (2015)

[29] Ashlea Alava Y. *et al.* High electron mobility and low noise quantum point contacts in an ultra-shallow all-epitaxial metal gate GaAs/Al$_x$Ga$_{1-x}$As heterostructure, *Appl. Phys. Lett.* 119, 063105 (2021)

[30] Kapfer M. *et al.* A Josephson relation for fractionally charged anyons, *Science* (2019) 363, 6429 pp. 846-849

[31] Serra, L. *et al.* Evidence for Majorana phases in the magnetoconductance of topological junctions based on two-dimensional electron gases, *Phys. Rev.* B 101 (11), 115409 (2020).

[32] Delfanazari K. *et al.*, Experimental evidence for topological phases in the magnetoconductance of 2DEG-based hybrid junctions, *arXiv*:2007.02057

[33] Shabani, J. *et al.* Two-dimensional epitaxial superconductor-semiconductor heterostructures: A platform for topological superconducting networks, *Phys. Rev. B* 93, 155402 (2016).

[34] Delfanazari K. *et al.* On-Chip Andreev Devices: hard Superconducting Gap and Quantum Transport in Ballistic Nb-In$_{0.75}$Ga$_{0.25}$As quantum well-Nb Josephson junctions, *Adv. Mater*., 29 1701836 (2017).

[35] Ren H. *et al.* Topological superconductivity in a phase-controlled Josephson junction, *Nature*, 10.1038/s41586-019-1148-9, (2019).

[36] Delfanazari K. *et al.* Proximity induced superconductivity in indium gallium arsenide quantum wells, *Journal of Magnetism and Magnetic Materials*, 10.1016/j.jmmm.2017.10.057, (2017).

[37] Hertel A. *et al.* Gate-Tunable transmon using selective-area-grown superconductor-semiconductor hybrid structures on silicon, *Phys. Rev. Applied* 18, 034042 (2022)

[38] Delfanazari K. *et al.*, Large-Scale On-Chip Integration of Gate-Voltage Addressable Hybrid Superconductor–Semiconductor Quantum Wells Field Effect Nano-Switch Arrays, *Adv. Electron. Mater. 2023, 2300453 DOI: 10.1002/aelm.202300453*.

[39] Delfanazari K. *et al.* On-chip Hybrid Superconducting-Semiconducting Quantum Circuit, *IEEE Transactions on Applied Superconductivity*, 28, 4, 1100304 (2018). 10.1109/TASC.2018.2812817,

[40] Aghaee M. *et al.* InAs-Al hybrid devices passing the topological gap protocol, *Phys. Rev. B* 107, 245423 (2023)

[41] Delfanazari K. *et al.* Quantized conductance in hybrid split-gate arrays of superconducting quantum point contacts with semiconducting two-dimensional electron systems. *Phys. Rev. Applied* 21.1 (2024): 014051.

*Appendix*

In Bauer's 1D tight-binding model, the interaction term in the Hamiltonian is given by *Eq*. (3.17) in Ref. 17, which we rewrite:

$$H_{\text{int}} = \sum_j U_j n_{j\uparrow} n_{j\downarrow}$$

(A1)

Here, $n_{j\uparrow(\downarrow)}$ counts for the number of electrons with spin ↑(↓) at site *j*; the on-site interaction strength $U_j = U$ is assumed to be constant at sites in the central constriction region and drops to zero at outer sites.

Based on *Eq*. (3.26) (though treated only to first order) in Ref. 1, which we rewrite here,

$$1 - U_{\text{eff}} = \frac{dV_c^h}{dV_c}$$

(A2)

1-$U_{\text{eff}}$ is equal to the differential of the effective Hartree barrier height $V_c^h$ with respect to the bare central barrier height $V_c$.

To acquire a DC bias spectroscopy, the drop of applied DC bias voltage $V_{DC}$ on series resistance $R_s$ should be removed. The corrected source-drain DC bias voltage $V_{SD}$ is

$$V_{SD} = V_{DC} - V_{DC} R_s \int_0^{V_{DC}} G_{SD} dV$$

(A3)

where, $G_{\text{DC}} = \int_0^{V_{\text{DC}}} G_{\text{SD}}\, dV$ is the DC conductance. We assume that the DC series resistance is equal to the AC series resistance, which is acquired by aligning the first plateau of $G_{\text{SD}}$ to $G_{\text{Q}}$.